# Blockchain Inspired Secure and Reliable Data Exchange Architecture for Cyber-Physical Healthcare System 4.0

Mohit Kumar[1], Hritu Raj[2], Nisha Chaurasia[1], Sukhpal Singh Gill[3]

[1]Department of Information Technology, National Institute of Technology Jalandhar Punjab India
[2]Department of Computer Science and Engineering, National Institute of Technology Jalandhar Punjab India
[3]School of Electronic Engineering and Computer Science, Queen Mary University of London, UK
kumarmohit@nitj.ac.in, hritur.cs.19@nitj.ac.in, chaurasian@nitj.ac.in, s.s.gill@qmul.ac.uk

**Corresponding author:** Mohit Kumar, Department of Information Technology, National Institute of Technology Jalandhar Punjab India -144011

## ABSTRACT

A cyber-physical system is considered to be a collection of strongly coupled communication systems and devices that poses numerous security trials in various industrial applications including healthcare. The security and privacy of patient data is still a big concern because healthcare data is sensitive and valuable, and it is most targeted over the internet. Moreover, from the industrial perspective, the cyber-physical system plays a crucial role in the exchange of data remotely using sensor nodes in distributed environments. In the healthcare industry, Blockchain technology offers a promising solution to resolve most securities-related issues due to its decentralized, immutability, and transparency properties. In this paper, a blockchain-inspired secure and reliable data exchange architecture is proposed in the cyber-physical healthcare industry 4.0. The proposed system uses the BigchainDB, Tendermint, Inter-Planetary-File-System (IPFS), MongoDB, and AES encryption algorithms to improve Healthcare 4.0. Furthermore, blockchain-enabled secure healthcare architecture for accessing and managing the records between Doctors and Patients is introduced. The development of a blockchain-based Electronic Healthcare Record (EHR) exchange system is purely patient-centric, which means the entire control of data is in the owner's hand which is backed by blockchain for security and privacy. The experimental results prove that the proposed architecture is robust to handle more security attacks and can recover the data if 2/3 of nodes are failed. The proposed model is patient-centric, and control of data is in the patient's hand to enhance security and privacy, even system administrators can't access data without user permission.

## 1. INTRODUCTION

In the growing world of technology, things around us becoming smarter than we think. The Sectors like healthcare industries are also revolutionaries by the latest technologies. As the technology grows, the Quality and efficiency of the healthcare industry are also increasing rapidly. Doctors and Patients both are getting the benefits of technological advancement in the healthcare industry. Now we are getting lab reports, MRI and CT scans in less time and are more efficient as well as accurate than earlier, Digital X-Ray revolutionaries the way look at fractures and tumors in bone, and digital storage of healthcare records opens a new way for patient care using deep learning and AI technologies. In addition, continuous remote monitoring of patients and collecting real-time data from the patients using IoT sensors, and performing the analysis without delay is possible due to advancements in technology [1]. We can now predict severe diseases (like cancer) more accurately and can prescribe medicine at a very earlier stage. Although storing healthcare data digitally offers many benefits but it also opens doors to security threats and data loss. As we know healthcare data is critical data, it consists of confidential and sensitive information related to patients. Hence, we require a reliable mechanism to ensure the integrity, security, and privacy of such types of sensitive data. Integration of blockchain technology with the healthcare industry can solve problems related to data integrity and security [2]. Now we can exchange health-related patient data more efficiently and securely with Doctors and healthcare providers.

Initially (in the 70's), the healthcare system is referred to as healthcare 1.0. There was a severe shortage of resources and restricted the ability to cooperate with digital systems in healthcare. Costs and time were both raised in the absence of embedded bio-medical sensors when healthcare companies turned to paper-based prescriptions and reporting during that period. The concept of healthcare systems began from 1991 to 2005 with healthcare 2.0. Digital tracking was used in this phase, enabling physicians to use imaging equipment to examine the health of patients. With the adoption of the internet platform, healthcare providers started to establish online communities and use cloud servers to store patient information which allowed for ubiquitous access for both the patient and the practitioner. Healthcare 3.0, gave rise to the concept of user-customization of patient healthcare records. New user interfaces enabled customized and optimized experiences. In addition to the advancements, healthcare record systems were implemented, which can track patients' medical data at the real-time, and universal level.

Similarly, stand-alone non-networked systems, such as social media channels, began to emerge alongside EHR systems, such as HL7, that were integrated to hold patient information. This reduced the sharing of health data, whether on the network or between clinicians using HL7. These methods also improved the ability to interact and communicate with patients. The Healthcare 4.0 era began in 2016 and will continue till the current day [3]. In this duration, a number of different technologies included fog computing, edge computing, cloud computing, the Internet of Things, advanced analytics, artificial intelligence, and machine learning, as well as blockchain to make it a smart healthcare system or Healthcare Industry 4.0. The primary focus was on wearable health sensors, so customized healthcare in real-time is possible. Figure 1 represents the illustrated view of the healthcare industry.

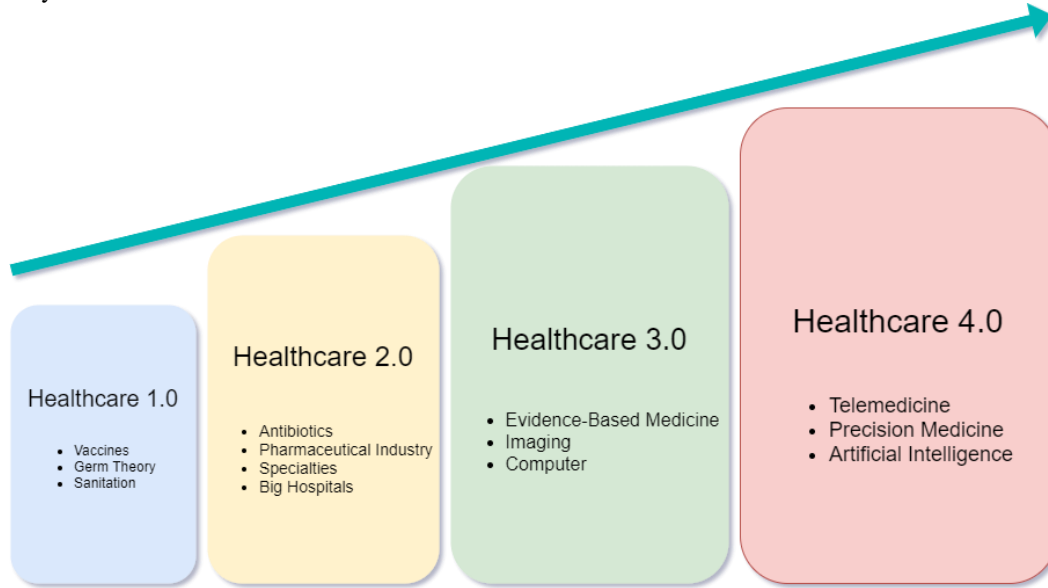

Fig. 1 Growth of the healthcare industry over time

1.1 Motivation: Blockchain becomes the prominent technology to ensure the security and privacy of user data and is used in several applications like healthcare, transportation, agriculture, smart home, supply chain, etc. The healthcare data of the patient is sensitive and valuable, and cannot be accessed by unauthorized users. The security and privacy of data is a challenging issue because healthcare data is the most targeted data over the internet. As the emergence of thrust technology especially blockchain along with IoT, Edge computing, and AI provides huge development in a smart healthcare system that was suffering from several issues like security attacks, reliability, centralized system, cost, latency, etc. The proposed healthcare system addressed the mentioned challenges and enhance the security, reliability, and transparency along with decentralized architecture. This research gives rise to blockchain-based solutions that will include privacy protection measures, data integrity, resistance to single-point-of-failure vulnerabilities, and safe information exchange among healthcare facilities. Also, the research benefits the healthcare industry stakeholders to better manage healthcare systems. It leads to the transformation from a traditional healthcare system (hospital-centric) towards a digital healthcare system (patient-centric). Also, this work helps to further scientific understanding, by enabling other researchers to further their research. The main contribution of the article is given as follows:

- We have proposed blockchain-enabled robust healthcare architecture for a patient-centric method that provides a cryptographic access control mechanism for a healthcare provider with a distinct medical institution.
- The proposed blockchain-based technique is used for creating a permission-based electronic health record-sharing system and investigating how the suggested system meets the requirements of patients, healthcare professionals, and others.
- The proposed blockchain model secures healthcare data from modern attacks and ensures integrity, a single point of failure, and authentication.
- The objective of the proposed approach is to provide a platform that is free from modern types of attack
- We finally decide the best course of action to optimize the blockchain system's performance based on the quality of service (QoS) parameters.

The remaining structure of the article is as follows: Section 2 gives a description of the related work toward the problem domain. The preliminary study towards the study is presented in Section 3. Section 4 represents the proposed system including advanced technology such as Interplanetary File System, BigchainDB, Tendermint, MongoDB, and cryptographic algorithms. The proposed blockchain-enabled secure architecture for the healthcare system is discussed in Section 5 and Section 6 dictates the implementation details, results, and discussion. At the last, the conclusion and future work is discussed in Section 7.

## 2. Literature Review

The healthcare system has been improved after the adoption of evolving technology such as the Internet of Things (IoT), Edge Computing, Artificial Intelligence, Machine learning, etc., but the security and privacy of healthcare-related data is still a challenging issue [36]. This consists of sensitive and confidential information about the patients and it could be risky for the patient's life if some data is tempered or leakage using modern attacks by attackers over the internet. Hence, researchers have proposed several approaches for managing and storing healthcare data. B. Shickel et al., provide a summary of deep learning-based approaches that has been applied for clinical applications to analysis EHR [6]. The authors have elaborated the limitations of existing framework, and models applied on heterogeneous data, patients record, and certificates. Basic image processing research is focused on more complicated and hierarchical representations of pictures and creative image processing with greater and more sophisticated structures. Z. Ying et al. proposed an efficient strategy for cloud computing environments using CP-ABE schemes, which maintain the CP-ABE for medical records exchange in a cloud environment [7]. The experimental results successfully demonstrated that the system can provide policy-preserving as well as attribute-recovery implementation with a little performance cost. X. Yang et al. published an article on sharing Electronic Health Records (HER) in the cloud with the help of blockchain technology [8]. They have designed a secure EHRs sharing system termed BVO-ABSC by using the blockchain and ABSC. The proposed model ensures the secrecy, accuracy, and unforgeability of EHRs. The cloud services are to authenticate the users' identification while preserving their privacy, and EHRs stored in the cloud. ABS enables the EHRs to be uploaded by authorized users and keeps the signers' identities anonymous, but cloud-based system have the high latency issue and cannot be used, where condition of patients is severe.

Y. Wang et al. dictate a method for exchanging medical records using a blockchain-enabled cloud computing approach [9]. Security and privacy are further accessed through authorization and other methods using blockchain. The basic design structure for consortium blockchain consists of the following critical components such as networking model, data creation, and consensus mechanism. The proposed approach was tested only Ethereum platform, and don't provide the guarantee to work with high efficiency at others platform also, storing and retrieving the data from cloud platform required the high bandwidth. Y. Zhuang et al., proposed a framework for exchanging information securely using blockchain technology and overcoming the barriers of the current system [10]. The authors introduce two modules: Request Module and Linkage Module. The developed model is not only ensured the privacy of patient data but also provides complete control of their healthcare data. The simulation results of the proposed framework proved that the proposed blockchain-based approach improves the current healthcare system in terms of QoS parameters like security, stability, and robustness. The main limitation of the proposed approach is that the performance of the system is depending upon the property of the blockchain node. Y. Yang et al., proposed a cloud-based patient data exchange system named MedShare that can deal with interoperability issues and overcome the barriers of the existing system [11]. Mediating the situation where independent healthcare providers are uninterested in sharing patient data with their consumers and showing a lack of desire in passing the data to their rivals. However, the reliability of proposed scheme is depending upon the public cloud and this approach required extra cost to implement data transformers.

F. Deng et al., introduced personal Health Record Exchange using Attribute-Based Signcryption [12]. They formally establish the CP-OABSC framework that relies on ABE schemes and verified out-sourced decryption but uses server-aided signature verification. A hybrid encryption technique is used that utilizes an attribute-based encryption approach and proves the system is accurate, secure, and robust. Z. Wu et al., [13] proposed a group-oriented bilinear pairing-based cryptosystem that can accommodate four encryption and decryption models. In order to have multiple keys, members only need to maintain one private key. A sender only has to conduct one round of encryption regardless of the models used. A. Ali et al., have proposed a patient-centric novel framework that guarantees the security and privacy of patient data while ensuring the minimum cost [23]. To enhance the security mechanism, authors have used a deep learning-based approach along with homomorphic encryption [24]. Further, group theory along with binary spring search (BSS) is applied to achieve trustworthiness, reliability, and confidentiality [25]. The whole idea was implemented using smart to reduce the anomaly in the IoT system. To access health records securely, multiple certificate authority (CA) along with Blockchain has been presented by the authors to overcome the limitations of single certificate authority [26]. To ensure the security of IoT devices, a lightweight authentication approach has been proposed that validates the latency and improves communication statistics and data preservation [27].

Two level security techniques have been proposed, where a blockchain-based approach is used to register and validate the patients using smart contract-based proof of work, further deep learning along with a Variational AutoEncoder (VAE) based approach is used to detect the intruder [28]. To verify the medical certificate and avoid the possibility of forging certificates of healthcare records was a challenging issue in the traditional system, now that has been addressed in Industry 4.0 with the help of thrust technologies like IoT, Blockchain, and AI [29-30]. The proposed guarantee to security solution with authentication and access control using smart contracts [31]. The application of blockchain in various domains such as security can share without any trusted third party [32], secure IoT infrastructure for advanced healthcare systems [33], and many more. Two techniques (node-based matrices and safeness scores) have been introduced for complex industrial systems by the authors to conceal the nodes from various Community detection algorithms [34]. The objective of the proposed approach was to minimize the persistence score and maximize the safeness score to achieve the desired result. The authors have introduced a privacy-preserving Distributed Application (DA) to generate, verify and maintain the healthcare-related medical certificates of patients using smart contracts [35]. Blockchain technology has been applied by researchers in diverse areas that are discussed in Table 1 along with techniques and limitations.

Table 1: Applications of Blockchain technique in diverse areas

| Year | Technique Used | Application | Limitations |
|------|----------------|-------------|-------------|
| 2018 [11] | MedShare | To extract the medical data, securely sharing and maintaining the patient data | Reliability is depending upon public cloud and required extra cost. |
| 2019 [14] | Dividing network participants into clusters and maintaining one copy of the ledger per cluster | Healthcare Data Management | Data can be compromised if the cluster head highjacked |
| 2019 [15] | Ethereum, IPFS Smart Contract | Storing (EHR) | Lack of control over data because some data is off-chain |
| 2020 [16] | Hyperledger Fabric Hyperledger Composer | Preserving Privacy of EHR (EHR) | Lack of use cases Complex architecture |
| 2018 [17] | Hyperledger Fabric | Healthcare | Complex architecture |
| 2019 [18] | N/A | Supply Chain Management | Only theoretical concept is given, no implementation details are available |
| 2017 [19] | Distributed Ledger Smart Contract | Finance | Lack of adoption because of decentralizing in nature |
| 2020 [20] | Hyperledger Fabric | Healthcare | Complex architecture |
| 2019 [21] | Smart Contract | EHR Management | Patient-Centric |

## 3. PRELIMINARY STUDY

**3.1 Security and Privacy Issues in Healthcare System:** Healthcare data of the patient is very sensitive that needs substantial security and privacy measures. The aim to maintain individual privacy of patient information is the beginning point when it comes to deciding who and whom should be granted access. To overcome this challenge, a number of security standards have been developed, including HIPAA, COBIT, and DISHA to protect patient data. The privacy of patients' information involves maintaining controls on access to the patient information, guarding against illegal access to patient data, and removing or destroying existing data. Figure 2 illustrates the hierarchical structure of security measures in Electronic Health Records.

**3.2 Statistics of Data breaches:** The U.S. Department of Health and Human Services (HHS) reported 3,687 health information thefts of 500 or even more records between 2010 and 2020 [4]. As a consequence, about 250 million healthcare records have been lost, stolen, exposed, or disclosed in error. As of 2018, there was around one data attack every day that involved 500 or even more records. When it hit Dec 2020, that rate increased by a factor of two. Figure 3 represents the graphical view of the number of breaches in years. Record exposure increased dramatically in 2015 and it was the worst in history for healthcare breaches due to the exposed, stolen, or impermissible disclosure of over 100 million data. Due to major cyber-attacks at three main health insurance providers Anthem, Premera Blue Cross, and Excellus, 2015 was a pretty horrible year. A better illustration is provided using the graph shown in Figure 4.

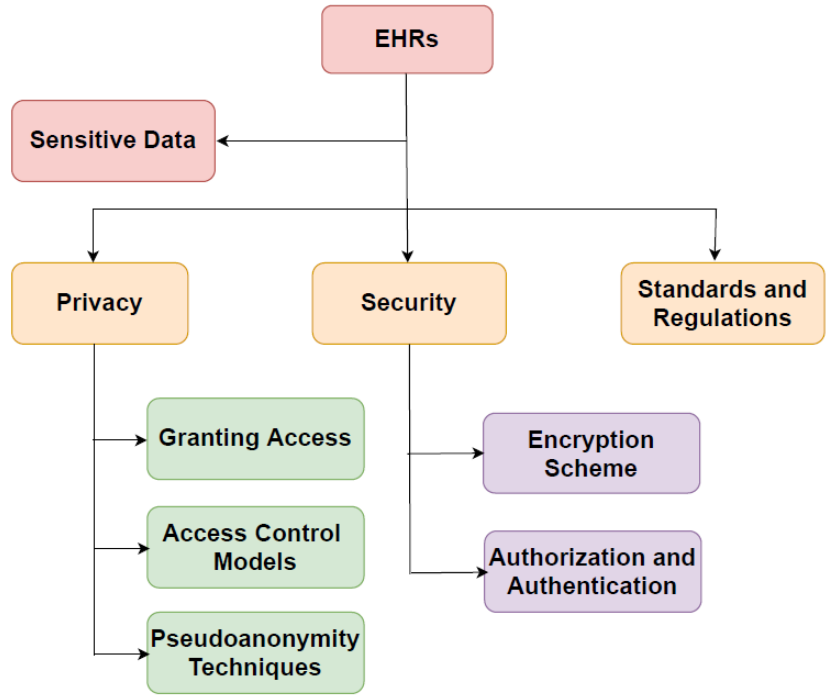

Fig. 2 Hierarchical structure for security and privacy measure in Cyber-Physical EHR

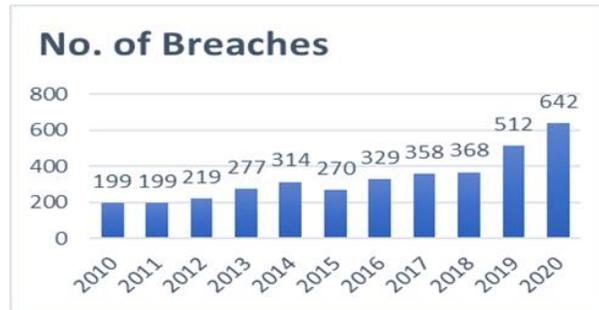

Fig. 3 No of Data Breaches Year-wise (2010-20)

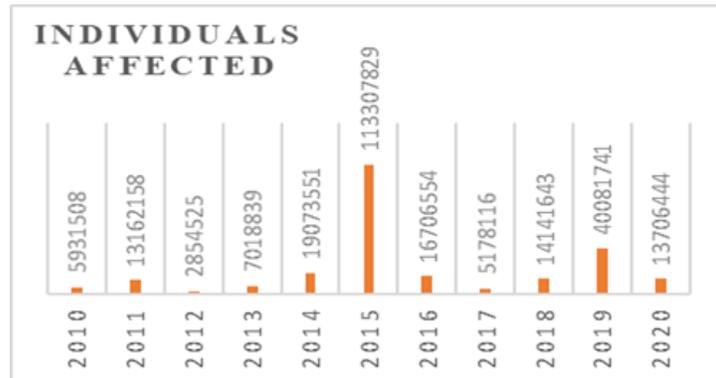

Fig 4: No. of Individuals affected each year (2010-20).

**3.3 Electronic Health Records:** Electronic Health Records (EHRs) are digital records that include information about a patient's medical history. A hospital or a clinician has the responsibility to maintain electronic medical records in a digital format throughout time. All relevant clinical data is essential for the treatment of the patient, including MRI reports, previous medical records, vaccinations, laboratory test results, and any patient allergies. This patient-specific information is easily accessible to the patient or the doctor and it is accessible only to authorized users. Sharing these results with only authorized care in the healthcare sector offers improvement in the research area.

**3.4 Blockchain:** A blockchain is a network of blocks that is interlinked with each other to save the records. An intriguing feature is that once information has been stored on a blockchain, it becomes impossible to alter. The hash of each block consists of a hash of the preceding block and the data that was in the block. For instance, let's consider an example, the chain of three blocks is represented in Figure 5, each block contains a hash and the hash of the previous block. the third block is connected to the second, and the second is connected to the first. The first one cannot connect to earlier blocks because it's the initial block and is named the genesis block. To put it another way, let's pretend you modified the contents of the second block. The resulting hash will change, too. Once the chain's data is invalidated, entire blocks are considered invalid since they cannot reference a valid hash of the preceding block. A change into a single block invalidates the blocks that follow it. But it is not enough to avoid manipulation by just hashing passwords.

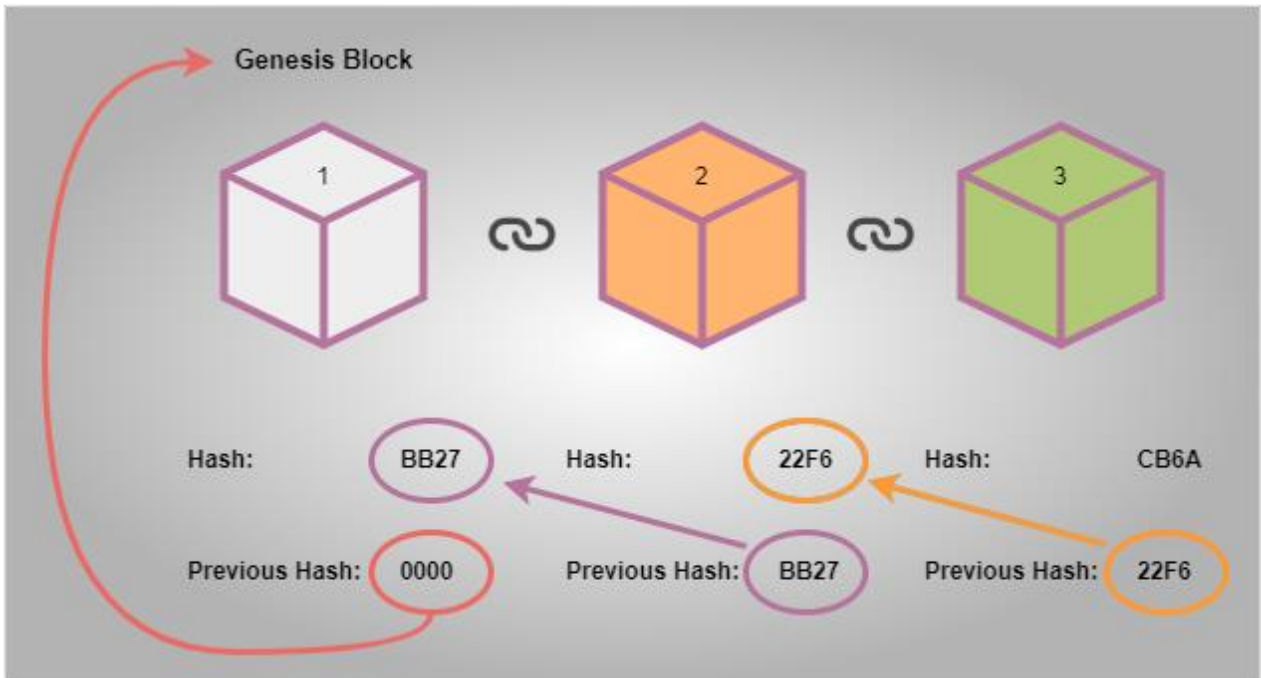

Fig. 5 Blockchain node with Genesis block and Hash values

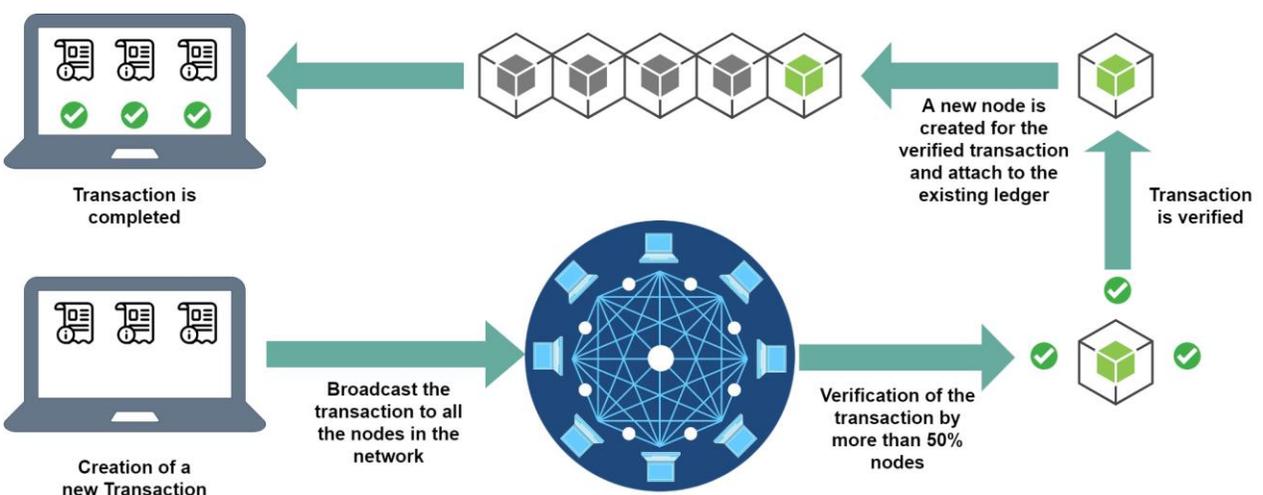

Fig. 6 Process of adding the new block in Cyber-Physical networks

In the world of cryptocurrency mining, computers have advanced computational power, they can do millions of hashes in one second. Recalculating all the hashes of subsequent blocks would fix your blockchain, too. To avoid this, blockchain uses the proof-of-work approach. This technique slows down the rate at which new blocks are created. It takes approximately 10 minutes for the proof-of-work calculation and includes a new block into the blockchain. This design makes it difficult to cheat by changing a single block because it required re-verifying the proof-of-work for all subsequent blocks. So, the security of a blockchain derives from its innovative use of hashing and proof-of-work, which is a way of distorting data to force it to match. Blockchain utilizes a peer-to-peer network rather than a central authority to govern the chain. Every new member receives the whole blockchain after joining the network. In this case, the node will utilize it to check whether everything is still intact. The process of adding a new block can be seen in Figure 6 where if a new record is created (called transaction), it is broadcasted to all the nodes in the network. Then at least greater than 50% of nodes validate it is a valid transaction. After that new block of corresponding transactions is added to the existing blockchain ledger and then the response of the successful transaction is received.

*3.4.1 Cryptographic Hash function*: It is a process that takes input and produces output, a one-way conversion [5]. Hashing is the process of turning the input of any length into a fixed-size string of text, using a mathematical formula. To put it another way, to find the hash of a message is to employ a function called a hash function, and the hash value is the result of that process. A cryptographic hash function must possess certain characteristics in order to be worthwhile. The output must have unique hashes. When an input is used to calculate the hash value, it implies that it should be impossible to de-rive the same hash value from several inputs, and, as a result, a single message should always provide the same hash value.

*3.4.2 Smart Contract*: In fact, smart contracts are more like standard contracts, which are used in the "real world." There is only one difference between smart and standard contact: everything is digital in smart contact. A smart contract is indeed a piece of software that is kept on a blockchain.

## 3.5 Proof of work:

**3.5.1 Preliminaries for the validation of transactions in Blockchain**: It is the responsibility of the blockchain to authorize legitimate entities and validate the transaction data using protocols. Further, it also ensures that the new node is a fake node or a legitimate node. A node does not trust the information it receives, so it performs a few checks using its own validation protocol during the process.

$$\mathcal{P}_r\mathcal{V}: \qquad \mathbb{N} * \mathcal{E} \rightarrow \{0, 1\} \qquad\qquad (1)$$

$$(\eta, \mathbb{T}_r) \rightarrow \mathcal{P}_r\mathcal{V}(\eta, \mathbb{T}_r) \qquad\qquad (2)$$

Where $\mathbb{N}$ is the set of nodes, $\mathbb{T}_r$ is the transaction and $\mathcal{E}$ represents the transactions set. Suppose a node is entered $\eta_{enter}$ in the network and computed $\mathcal{P}_r\mathcal{V}(\eta_{enter}, \mathbb{T}_{rX})$ for the validation. If the value is 1 then the transaction is valid and broadcasted to the nearby nodes otherwise it is rejected. Transaction $\mathbb{T}_{rX}$ eventually reaches a complete node $(\eta_c)$ that verifies the identity of the sender after entering into the network. If $\mathbb{T}_{rX}$ is valid then $\eta_c$ append it with a list of valid transactions using equation 3

$$\mathcal{L}_l^{\eta_c}.\text{append}(\mathbb{T}_{rX}) \qquad\qquad (3)$$

Eventually, a new block (representing a set of valid transactions) is created out of a subset of transactions in the local list of $\mathcal{L}_l^{\eta_c}$:

$$\mathcal{B}_{\eta_c} = (\mathbb{T}_r^{1}, \mathbb{T}_r^{2} \ldots \ldots \ldots \mathbb{T}_r^{N}) \qquad\qquad (4)$$

Where $\mathbb{T}_{r_i} \in \mathcal{L}_l^{\eta_c}$ and $\mathcal{B}_{last}^{\eta_c}$ represents the last block in the local chain of $\eta_c$, according to protocol and it starts to solve the proof-of-work for $(\mathcal{B}_{last}^{\eta_c}, \mathcal{B}_{\eta_c})$. Many $\eta_c$ are doing this parallelly i.e., competing with each other.

After a given time frame, the node generates a new block and starts solving for the next block. Suppose a new block $\mathcal{B}_{new}$ along with its header is created by node n and sent to the nearest node in the network. $\mathcal{B}_{new}$ is received by a node $\eta_r$, if any of the transactions $\mathbb{T}_r$ in $\mathcal{B}_{new}$ is found invalid then the whole block is rejected. A complete node $\eta_c$ must decide whether to add $\mathcal{B}_{new}$ to its local blockchain. If $\mathcal{B}_{new}$ is added to the local version of its blockchain, then a hash of the previous block ($\mathcal{B}_{prev}$) in the transmitted header Head ($\mathcal{B}_{new}$) and the hash of the last block in the local blockchain are equal. Since $\eta_c$ has been accepted $\mathcal{B}_{new}$, it removes the set of transactions present in this block from its local list. Suppose at the time $t_0$, entire nodes of networks are agreed to the same version of the blockchain, say $\mathcal{B}k_0$ and let $\mathcal{B}_1$ and $\mathcal{B}_2$ be two blocks broadcasted by two different nodes at some time $t_1$, where $t_1 > t_2$. There might be some time delay for blocks to reach some nodes, a condition may arise when some nodes receive $\mathcal{B}_1$ before $\mathcal{B}_2$. Some nodes will have their local version of the blockchain as given in equations 5 & 6.

$$\mathcal{B}k_0 + \mathcal{B}_1 \qquad\qquad (5)$$

$$\mathcal{B}k_0 + \mathcal{B}_2 \qquad\qquad (6)$$

The rule of consensus that forms the security of blockchain describes that a node will always keep the version of the chain with the largest work. Let's assume that $t_3 > t_2$, one of the nodes with the version of local blockchain $\mathcal{B}k_0 + \mathcal{B}_1$ emits a Block $\mathcal{B}_2$ into the network i.e., the local blockchain for this node will look like $\mathcal{B}k_0 + \mathcal{B}_1 + \mathcal{B}_3$. When the broadcasted block $\mathcal{B}_3$ reaches a node (say $\eta_f$) with local blockchain $\mathcal{B}k_0 + \mathcal{B}_2$, there will be a problem. Since $\mathcal{B}_3$ needs $\mathcal{B}_1$ as the previous block, it can't be added to the local chain of $\eta_f$ (dilemma condition). Since the dilemma is for the blocks after $\mathcal{B}k_0$, there are two options: $\mathcal{B}k_0 + \mathcal{B}_1 + \mathcal{B}_3$ and $\mathcal{B}k_0 + \mathcal{B}_2$. As per protocol, the longest chain will be kept i.e., $\mathcal{B}k_0 + \mathcal{B}_1 + \mathcal{B}_3$ and the transaction of block $\mathcal{B}_2$ is again taken into consideration for the new block. As the depth of the chain goes on increasing, the confidence for the deeper nodes keeps on increasing in the whole blockchain network. Ultimately all the nodes in the network achieve the consensus.

## 4. PROPOSED MODEL

The proposed blockchain-enabled cyber physical EHR sharing system architecture uses different techniques and systems to manage the block transactions. The system's EHR may be disseminated to blockchain network members via a shared symmetric key and public key. A simple stepwise process of the proposed system is shown in Figure 7 which consists of several entities like the patient, doctor, blockchain, and others. Internet of Things (IoT) devices or sensors are used to collect healthcare data from patients and send it to edge devices for processing without delay. The entire records of patients' data are saved at edge devices by developed applications.

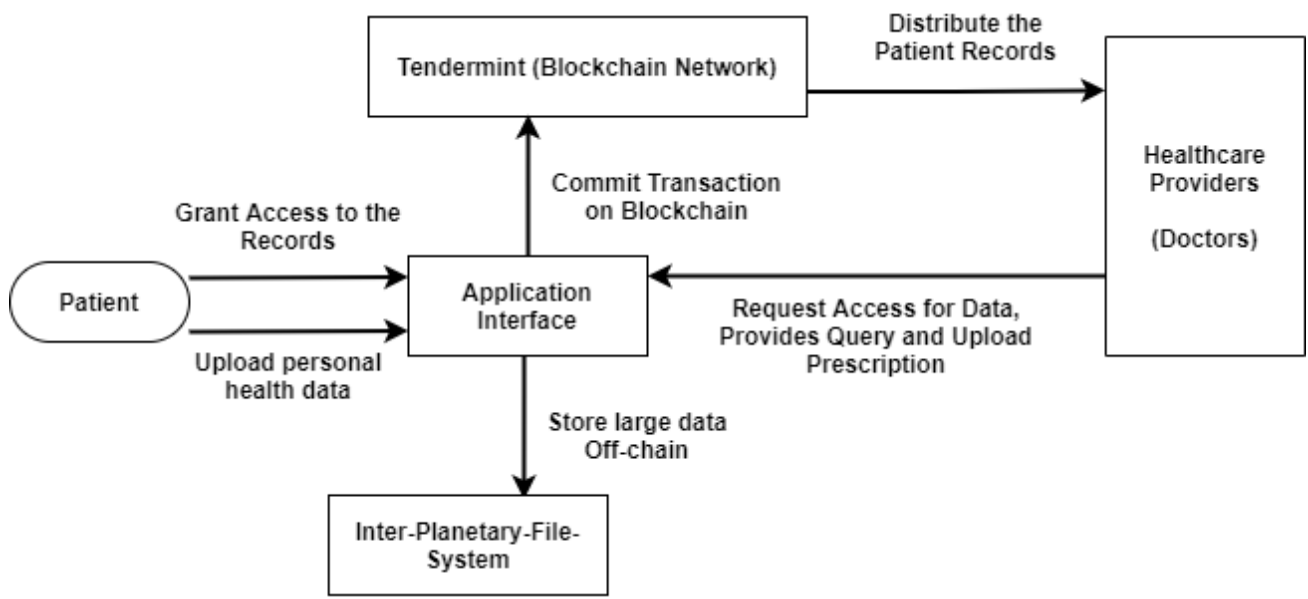

Fig. 7 Securing process of health records in the Cyber-Physical blockchain environment

The registered doctor sends the request for accessing that data and the request is granted only to legitimate doctors, and allows to upload the prescription for the patients based on the healthcare data. In addition, doctors cannot share the sensitive data of patients without permission. Here, blockchain technology plays an important role to secure the data over the internet during transmission. It is the responsibility of the blockchain to authorize legitimate entities and validate the transaction data using protocols. The basic functionalities of different tools and technologies used in the implementation of blockchain-based EHR are discussed briefly.

**A. Interplanetary File System (IPFS):** The IPFS file system stores and shares data by implementing a peer-to-peer (P2P) system. For every item of material that is uploaded to IPFS, a unique hash is associated with the item. Even if you make only one single change, the hash stays completely unique. Instead of utilizing a domain name, IPFS may utilize the content of the file to determine its location, and the data's name is unchangeable. Entire data references are kept in a Kademlia-based DHT. Routing involves announcing fresh data to the network and also helps to find requested data. A number of small data values (about 1KB in size) are directly integrated into the DHT. As the DHT becomes large, it stores references of the node that have the block data.

**B. BigchainDB:** It is Built on High Throughput, Low Latency, Powerful Query Capabilities, Decentralized Control, Immutable data Storage, and Built-in Asset Support. Because the software is open-source, it supports several programming languages (Java, Python, and Javascript), as well as Docker. These transactions are defined in a real-time database. The initial CREATE transactions enable users to create new records in the database. The second transaction is the TRANSFER transaction, which transfers ownership of a specified record to another user. Another important aspect of the BigchainDB transaction blocks is that they consist of three major elements: Asset, Metadata, and Transaction ID.

**C. Tendermint:** It is a piece of software that may be used to reliably and securely replicate a program on many computers. Tendermint continues to operate even if one-third of computers may potentially fail in random ways. The faulty machines never view the same transaction log, and the states computed by each computer are always different. A basic issue in distributed systems is the lack of reliable and consistent replication; it occurs in many applications such as currencies, elections, and infrastructure orchestration, and it is essential for system fault tolerance. If a system can stand with malfunctions, such as turning malevolent occurred is called Byzantine fault tolerance (BFT). The block diagram of the tendermint node is shown in Figure 8.

**D. MongoDB:** A username and password may be kept in an authentication system, but in the case of blockchain or IPFS, using these methods on the user interface carries a cost burden. As a result, all additional local information was stored in a replacement storage location, and each account was assigned a cryptographically secure public key. Thus, MongoDB was the logical selection.

**E. Cryptographic Algorithm:** To restrict access to the data and allow some specific people, it is necessary to use some kind of encryption. Symmetric encryption includes a key used to encrypt documents and convert them to unreadable form. In order to recover the original document, the decryption process is required using the same key. We have used Advanced Encryption Standard (AES) as their symmetric algorithm for encryption and decryption to implement the blockchain-enabled proposed architecture for the healthcare system.

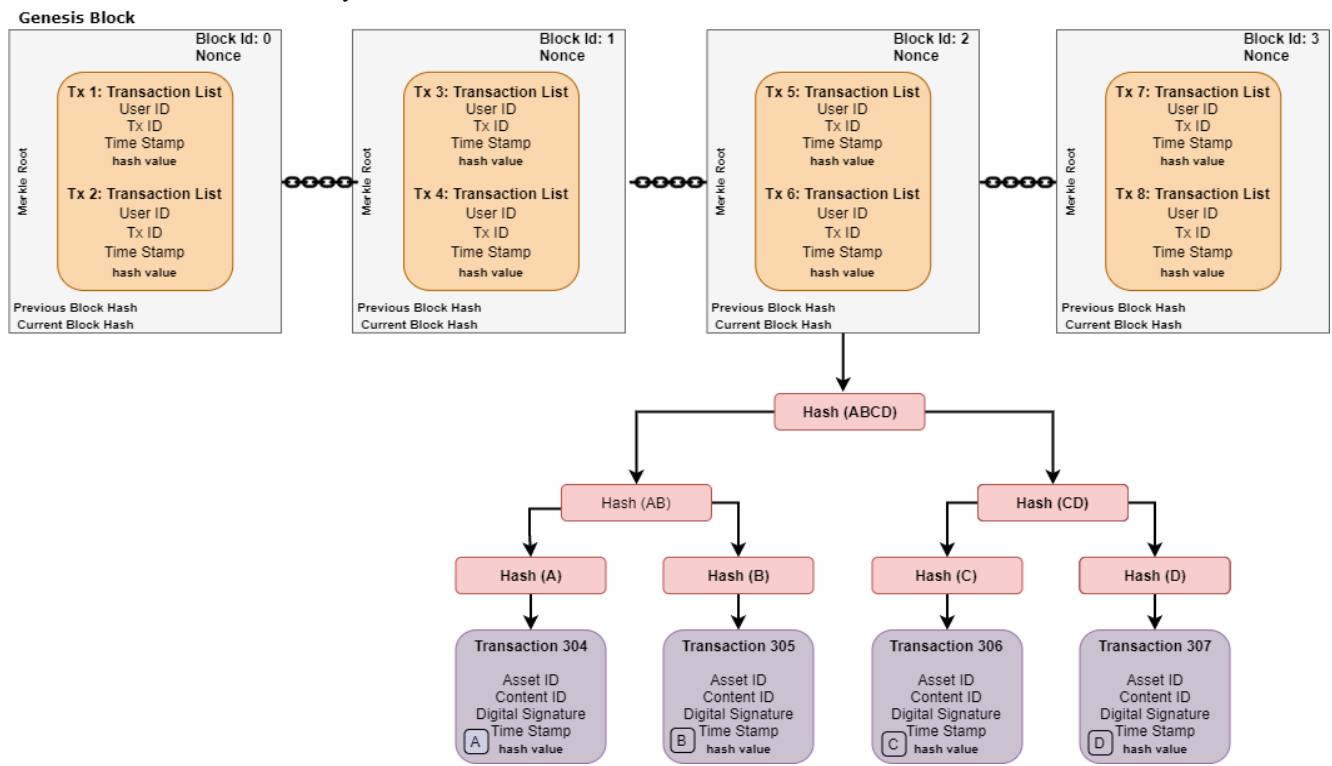

Fig. 8 Block diagram of data in Tendermint node

## 5. Blockchain-enabled Secure Architecture for Healthcare Systems

We have proposed blockchain-enabled secure architecture for the healthcare system that is patient-centric and used for creating a permission-based EHR sharing system. The aim of the proposed architecture is to provide a platform that is free from modern types of attack. The proposed architecture shown in Figure 9 demonstrates the deployment of blockchain technology for exchanging EHR in three levels. Different IoT sensors have been used to collect health-related data like Heart Rate, Blood Pressor, Motion Sensor, etc., which is attached to the Patient's body for various purposes like real-time monitoring of heart rate, and oxygen saturation level for enhanced treatment. Some of these are implanted devices that have resource constraints especially unable to perform computation. Hence, an emerging computing paradigm named edge devices are used to send the data or process a limited amount of time-sensitive data. The data is stored, shared, and updates by the entities of healthcare personnel in the blockchain networks. This is a medical software application that allows doctors and patients to transmit and receive health information or services from distant locations using blockchain technology. The proposed blockchain-enabled architecture ensures the privacy and security of patient data.

## 6. IMPLEMENTATION DETAILS, RESULTS, AND DISCUSSION

This section has been divided into two sections. The first section dictates the experimental setup for implementing the proposed blockchain approach, the AES algorithm, storage of encrypted files and details about the blockchain transaction. The second section discusses the registration of patients and doctors at the developed platform, adding the records of patients, collection of patient data, and securing the healthcare data of patients and results in detail.

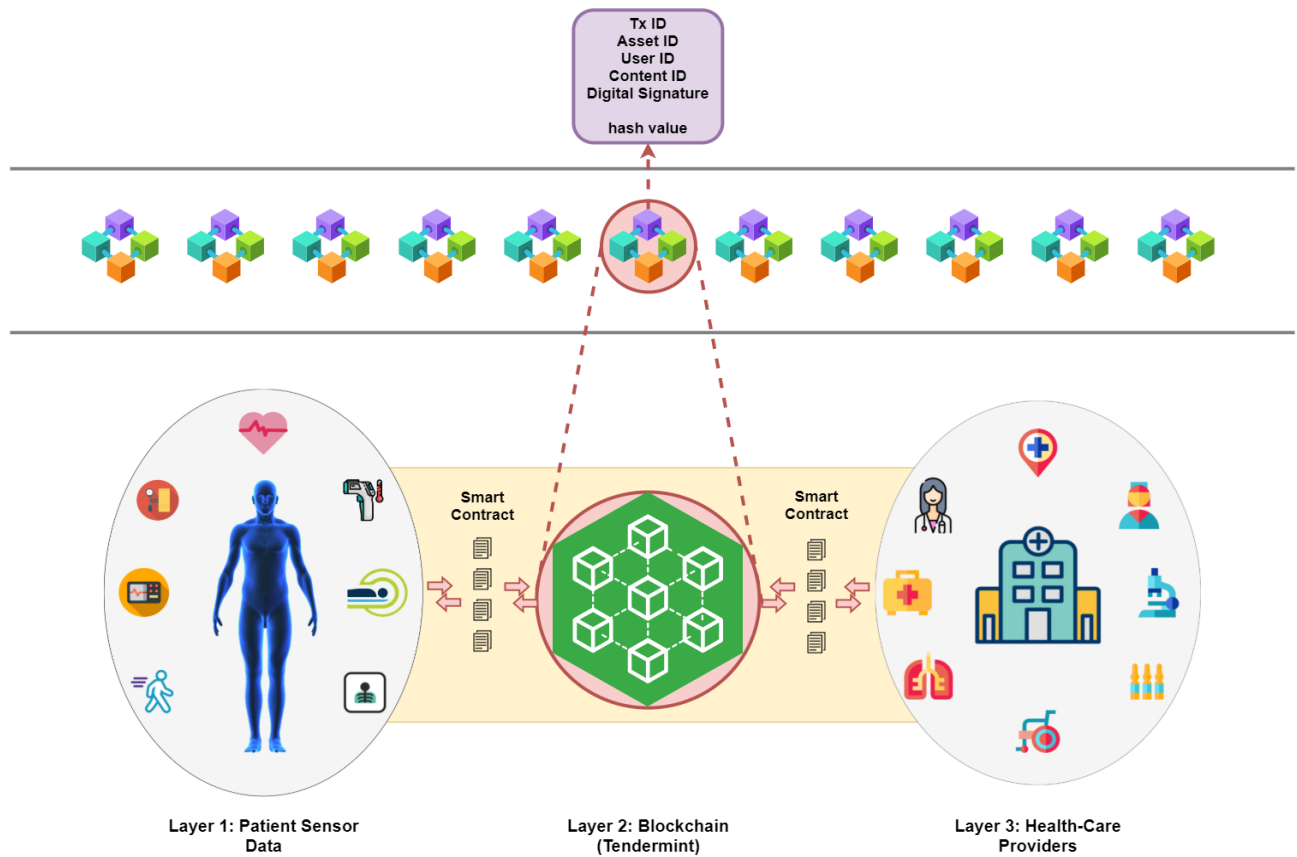

Fig.9 Blockchain-based proposed Cyber-Physical Architecture

**6.1 Implementation Details:** Medical data may be safely kept in an open database like the IPFS by using the aforementioned technologies. The massive data is not kept on the blockchain itself, but on the distributed network that uses the IPFS, to make this as a function, and start the procedure. Healthcare 4.0, combined with Tendermint blockchain and the Interplanetary File System (IPFS), can introduce transformative changes to the healthcare industry. Blockchain can play a significant role by providing a secure, transparent, and decentralized infrastructure for various healthcare applications. Let's discuss some technical aspects of how Tendermint blockchain can be applied in healthcare 4.0:

**Secure and Immutable Data Storage:** Blockchain, in conjunction with IPFS, can provide a secure and decentralized infrastructure for storing healthcare data. IPFS allows for the distributed storage of files, utilizing a content-addressable system where files are identified by their unique hash. This ensures data integrity and prevents tampering or unauthorized modifications. By storing the file hashes on the blockchain, the immutability and transparency of data storage can be ensured.

**Data Interoperability and Exchange:** One of the key challenges in healthcare is the fragmented nature of patient data across different healthcare providers and systems. Blockchain can facilitate data interoperability and exchange by creating a unified and standardized platform for securely storing and sharing patient health records. Using a decentralized blockchain, patients can have control over their data and grant permissions to healthcare providers or researchers, ensuring data privacy and consent management.

**Secure Medical Records:** Blockchain can be used to create an immutable and tamper-resistant ledger for storing electronic health records (EHRs) and medical information. Each transaction or update to the medical records can be recorded as a block on the blockchain, ensuring transparency, data integrity, and protection against unauthorized modifications. This secure and auditable storage mechanism enhances patient trust and enables seamless access to medical records across different healthcare providers.

**Consent Management and Privacy:** Patient consent management is critical in healthcare, particularly when sharing sensitive medical information for research or treatment purposes. With blockchain, consent management can be implemented using smart contracts. Patients can define granular consent rules and grant temporary access to their medical data for specific purposes or periods. The blockchain's transparent nature ensures that consent agreements are recorded, immutable, and enforceable, enhancing privacy and trust between patients and healthcare stakeholders.

**Clinical Trials and Research:** Blockchain can streamline and enhance the integrity of clinical trials and medical research. By recording the entire research process, including protocols, data collection, and analysis, on the blockchain, transparency and traceability are ensured. This enables verifiable and reproducible research results, reduces fraud, and improves the overall credibility of clinical trials. Additionally, blockchain-based incentive mechanisms, such as tokenization or rewards, can encourage patient participation in trials and data sharing.

**Telemedicine and Remote Patient Monitoring:** With the increasing adoption of telemedicine and remote patient monitoring, Blockchain can provide a secure and transparent infrastructure for managing patient data and ensuring the integrity of remote healthcare services. Blockchain-based smart contracts can facilitate automated payments, enforce service-level agreements, and maintain an auditable log of telemedicine interactions and patient monitoring data.

**Scalability and Performance:** Blockchain's consensus algorithm and IPFS's distributed file storage enable scalability and high-performance data handling. Tendermint's PBFT-based consensus algorithm allows for fast transaction finality and higher throughput, suitable for managing a large volume of healthcare data. IPFS's distributed storage architecture facilitates parallel retrieval and storage of files, ensuring efficient data access and retrieval.

It's important to note that while blockchain offers several advantages for healthcare applications, implementation considerations, regulatory compliance, scalability, and integration with existing systems are key challenges that need to be addressed. Realizing the full potential of healthcare 4.0 requires collaboration among healthcare providers, technology experts, regulators, and policymakers to design and deploy robust and interoperable blockchain solutions.

**6.1.1 Key management**: In Tendermint blockchain, keys play a crucial role in establishing the identity and permissions of network participants, particularly validators. Here's a detailed explanation of how keys are managed in Tendermint:

**Key Generation:** The process starts with the generation of cryptographic key pairs by network participants. The key pair consists of a public key and a corresponding private key. The private key must be securely stored and kept secret by the owner, while the public key is shared with the network.

**Validator Identity:** In Tendermint, validators are identified by their public keys. When joining the network, a validator's public key is registered and associated with their identity. This ensures that validators can be uniquely identified and authenticated during the consensus process.

**Consensus Signing:** Validators in Tendermint use their private keys to digitally sign messages during the consensus protocol. These signatures provide cryptographic proof of the validator's authenticity and ensure the integrity and validity of consensus-related messages exchanged among validators.

**Authentication:** Keys are used for authentication purposes within the blockchain network. Validators use their private keys to sign and verify messages, establishing their identity and proving that they are authorized participants in the consensus process. This authentication process helps prevent unauthorized actors from participating in the consensus and protects the network from malicious activity.

**Secure Key Storage:** As private keys are essential for signing messages and asserting identity, it is crucial to securely store them. Validators must safeguard their private keys to prevent unauthorized access and potential compromise. Common practices include storing private keys in hardware security modules (HSMs), encrypted key stores, or using secure offline storage solutions.

**Key Rotation:** To enhance security, It encourages regular key rotation. Validators can periodically generate new key pairs and associate the new public keys with their identity. This process helps mitigate the risk of key compromise and strengthens the overall security of the network.

**Key Management Systems:** In real-world deployments, key management systems (KMS) may be employed to enhance key security and management. KMS solutions provide centralized control and protection of keys, allowing for key generation, rotation, storage, and access control. They can integrate with Blockchain network to streamline key management processes and enforce best security practices.

Overall, key management is crucial in Tendermint blockchain to establish and verify the identities of network participants, ensure the integrity of consensus messages, and protect the network from unauthorized access. Proper key generation, storage, rotation, and management practices are essential to maintaining a secure and trusted blockchain network.

**6.1.2 Involvement of entities to improvement the performance**: Several entities are involved in Tendermint for performance improvement. Here are some key entities and their roles:

**Validators:** Validators are the network nodes responsible for participating in the consensus process and validating transactions. They propose new blocks, vote on the validity of proposed blocks, and participate in the consensus protocol to agree on the state of the blockchain. Validators play a crucial role in improving performance by efficiently processing and validating transactions.

**Consensus Algorithm:** It utilizes a consensus algorithm based on Practical Byzantine Fault Tolerance (PBFT). This algorithm enables validators to reach agreement on the order and validity of transactions in a distributed network. The consensus algorithm ensures that the network can achieve high throughput and low latency, leading to improved performance.

**Block Propagation:** It employs a gossip-based protocol for block propagation. When a validator proposes a new block, it disseminates the block to a subset of other validators, who then further propagate it throughout the network. This approach helps in reducing the propagation time of blocks, enhancing performance by minimizing the delay in reaching consensus.

**Block Time and Finality:** It allows for the configuration of block time, which determines how frequently new blocks are added to the blockchain. By adjusting the block time, network operators can optimize performance according to the requirements of their specific use case. Additionally, It provides fast finality, meaning that once a block is committed, it is considered finalized, reducing the need for lengthy confirmation times.

**Peer-to-Peer Networking:** It utilizes a peer-to-peer networking layer to facilitate communication among nodes in the network. It optimizes network performance by efficiently transmitting blocks and messages between validators. The networking layer employs various techniques, such as protocol buffers and efficient data serialization, to enhance the speed and reliability of data transmission.

**Parallel Processing:** It allows for parallel processing of transactions. This means that validators can process multiple transactions concurrently, improving overall throughput and performance.

By leveraging these entities and techniques, Tendermint aims to achieve high-performance blockchain networks with fast transaction processing, low latency, and scalability. These performance improvements are crucial for building decentralized applications and supporting many users and transactions on the blockchain.

**6.1.3 Security Scenario:** When a person's health data leaks, it can pose significant security risks. Here's a detailed explanation of the potential risks associated with health data breaches:

**Identity Theft:** If an individual's health data, including personally identifiable information (PII) such as name, address, social security number, or medical insurance details, is exposed, it becomes easier for malicious actors to commit identity theft. They can use the stolen information to create fraudulent accounts, make unauthorized transactions, or access other sensitive personal and financial data.

**Medical Identity Theft:** Health data breaches can lead to medical identity theft, where an unauthorized person uses someone else's health information for personal gain. This can result in fraudulent medical billing, obtaining prescription drugs illegally, or receiving medical treatment using the victim's identity, potentially leading to incorrect diagnoses or inappropriate medical care.

**Financial Consequences:** If health data breaches expose financial information, such as credit card details or banking information, individuals can face financial consequences. Cybercriminals may use the stolen data for fraudulent transactions, draining bank accounts, or making unauthorized purchases, leading to financial losses and potential damage to credit scores.

**Stigmatization and Discrimination:** Certain health conditions, such as mental health disorders or sexually transmitted diseases, may carry social stigma or lead to discrimination. If such sensitive health data is leaked, individuals may face discrimination in various areas of life, including employment, insurance coverage, or personal relationships.

**Medical Fraud and Insurance Abuse:** Compromised health data can be exploited for medical fraud or insurance abuse. Criminals may use the stolen information to submit false insurance claims, obtain prescription medications for illegal resale, or fraudulently bill for medical services not provided. This type of fraud can result in financial losses for insurance companies, healthcare providers, and individuals themselves.

**Reputational Damage:** When health data leaks, it can cause significant reputational damage to individuals and healthcare organizations. The loss of trust from patients or customers can have long-lasting consequences for healthcare providers, leading to a decline in patient loyalty, negative publicity, and potential legal actions.

**Targeted Attacks and Spear Phishing:** Health data breaches can provide attackers with valuable information to launch targeted attacks, such as spear phishing. With knowledge of an individual's health conditions or medical history, cybercriminals can craft convincing and personalized phishing emails, aiming to deceive the victims into providing more sensitive information, financial details, or login credentials.

**Research and Intellectual Property Theft:** If research data or intellectual property related to healthcare innovations or drug development is exposed, it can lead to intellectual property theft and compromise ongoing research efforts. This can result in financial losses for research organizations, setbacks in medical advancements, and potential harm to public health.

### A. Environment Initialization

- Start the BigchainDB server and connect it to MongoDB running on localhost.
- Configure the tendermint to connect with BigchainDB then run the tendermint core.
- Start Block Dashboard API on localhost:3000 to view the blockchain blocks
- Start the Node.Js server and run the app.

**B. Configuration of System:** The configuration of the system is shown in Table 2 for the implementation of the proposed blockchain-enabled healthcare architecture.

Table 2: System Configurations

| Resource | Specialization |
|---|---|
| OS | Ubuntu 21.04 |
| Processor | Intel(R) Core(TM) i7-4500U CPU @ 1.80GHz |
| RAM | 8.00 GB |
| Programming Language and API | Java Script (React Library and Express Js) and Node Js |

**C. Generate Private and Public Keys:** EdDSA generates the private and public keys for everyone in the network. The private key is securely kept, while the public key is available for all people. Every participant publicly disclosed the public key, but they have the private key corresponding to it [22].

**D. Data encryption using AES algorithm**: We have used the AES algorithm for EdDSA key and Signature generation as shown in algorithm 1. This data will be encrypted using the AES-256 symmetric encryption method. The first step is to get a key that is created randomly. This random key is used to encrypt the report and produce an encrypted file. This key can only be created from scratch if you first encrypted the file. The encryption and decryption function is shown in Figure 10.

Algorithm 1. EdDSA key and Signature generation

Input: random key, public key.
Output: EdDSA key and Signature
Key Setup.
1: Hash k such that $H(k) = (h_0, h_1, \ldots\ldots\ldots h_{2b-1})$
2: $a = (h_0, h_1, \ldots\ldots\ldots h_{b-1}$ interpret as integer in little-endian notation
3: $b = (h_b, \ldots\ldots\ldots h_{2b-1})$.
4: Compute the public key $A = aB$.
   Generate the signature
   a. Calculate the ephemeral key: $r = H(b, M)$.
   b. Calculate the ephemeral public key: $R = rB$.
   c. Calculate the $h = H(R, A, M)$ and convert it to integer.
   d. Calculate the $S = (r + ha) \mod l$.
   e. Signature pair: (R, S).
Return output

```
function encrypt(text){
    var cipher = crypto.createCipher('aes-256-cbc','d6F3Efeq')
    var crypted = cipher.update(text,'utf8','hex')
    crypted += cipher.final('hex');
    return crypted;
}
function decrypt(text){
    var decipher = crypto.createDecipher('aes-256-cbc','d6F3Efeq')
    var dec = decipher.update(text,'hex','utf8')
    dec += decipher.final('utf8');
    return dec;
}
```

Fig. 10 Encryption and decryption function using AES-256

**E. Store the file encrypted using AES on IPFS:** The data will be stored on IPFS. Once a record is saved on IPFS, it produces a unique hash. The hash key grants access to the data. It has now been made available on the blockchain as shown in Figure 11.

Fig. 11 Function to store file on IPFS

**F. Commit the transaction on Blockchain:** The transaction will now commit in BigchainDB on the local node as well as on the peer node using Tendermint as shown in Figure 12.

Fig. 12 Committing the transaction on BigchainDB.

**G. Granting and Revoking access to the doctor:** As we discussed earlier, this is a patient-centric application. So, the patient has all the control of their data. They can grant access and can revoke access to or from the Doctor. We have implemented a function for this, where patients grant access to the doctor and patients can revoke to the doctors.

**H. Asset retrieval**: Figure 13 shows the complete flow to retrieve the assets which is stored in BigchainDB and IPFS. To retrieve the data, start the node.js server, connect it to the IPFS network then connect to BigchainDB. After a successful connection calls the getAssetId function with the respective user public key.

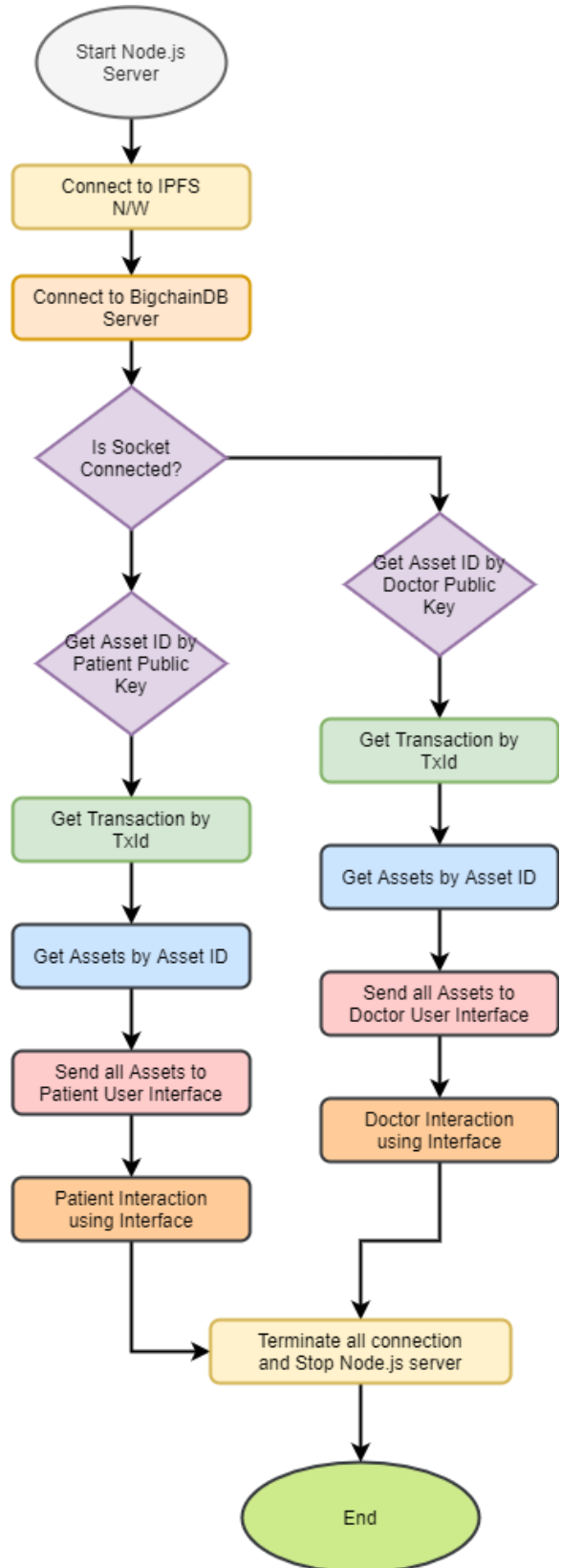

Fig. 13 Backend view of asset retrieval

**I. Front-end view of Doctor and Patient**: The Patient's view and Doctor's view is shown in Figure 14 and 15 respectively. We will now address the individual in question and provide the necessary procedures for keeping his healthcare data in MediChain, which medical personnel may seek and access if allowed. Figure 16 illustrates a rough sketch of these processes.

- Registered patients can log into the portal.

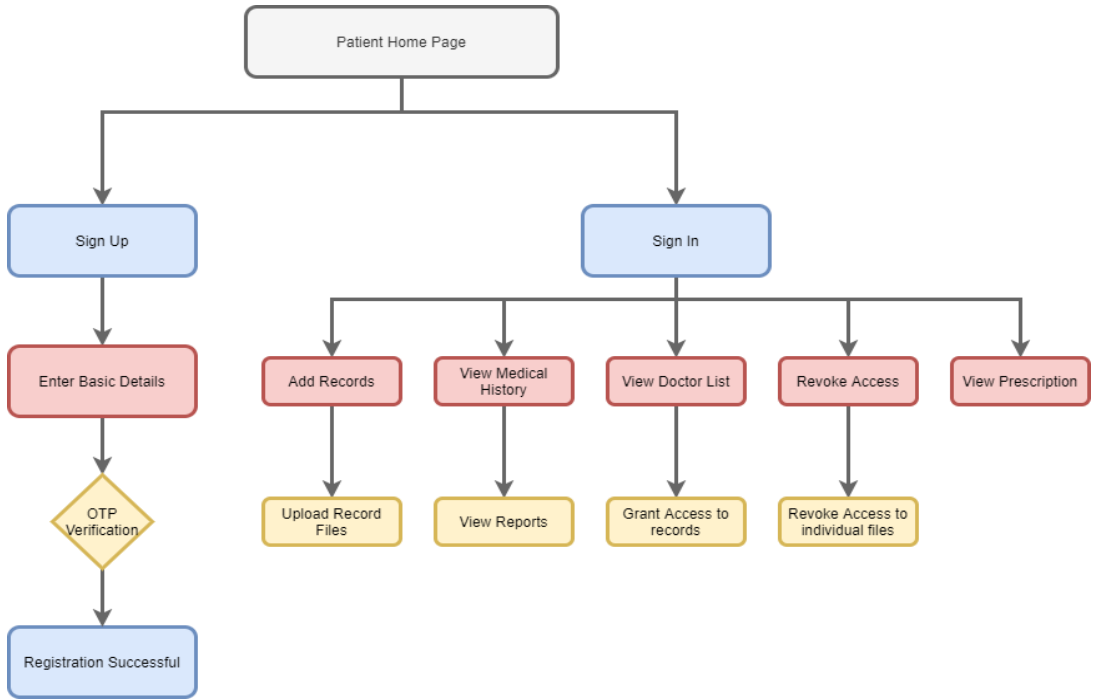

Fig. 14 patient view

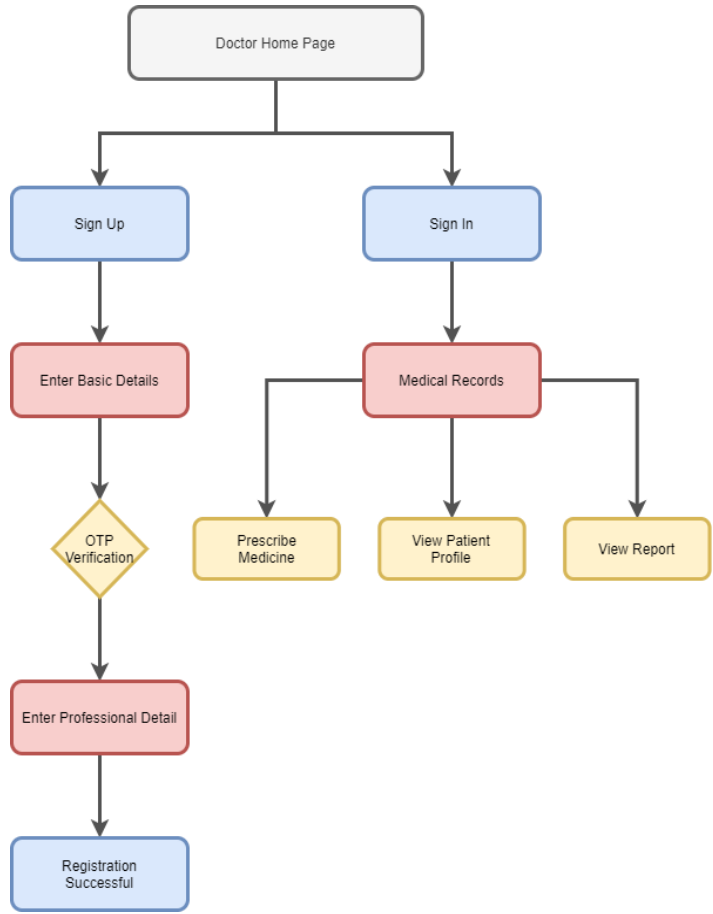

Fig. 15 Doctor's view

- A patient can upload his/her past medical history and other health-related information.
- In the other hand, doctors and other health-related persons can log in to their respective portals.
- They can upload clinical data.
- Request patients for granted access to their health-related data.
- Patients can allow access to their data if they want.
- After getting access to data, then doctors can prescribe accordingly.
- Patients can also revoke access to their data anytime if they want.

All large files like Radiology scans and large lab reports are pushed to off-chain (IPFS) and encrypted Content ID (CID) are stored on the blockchain.

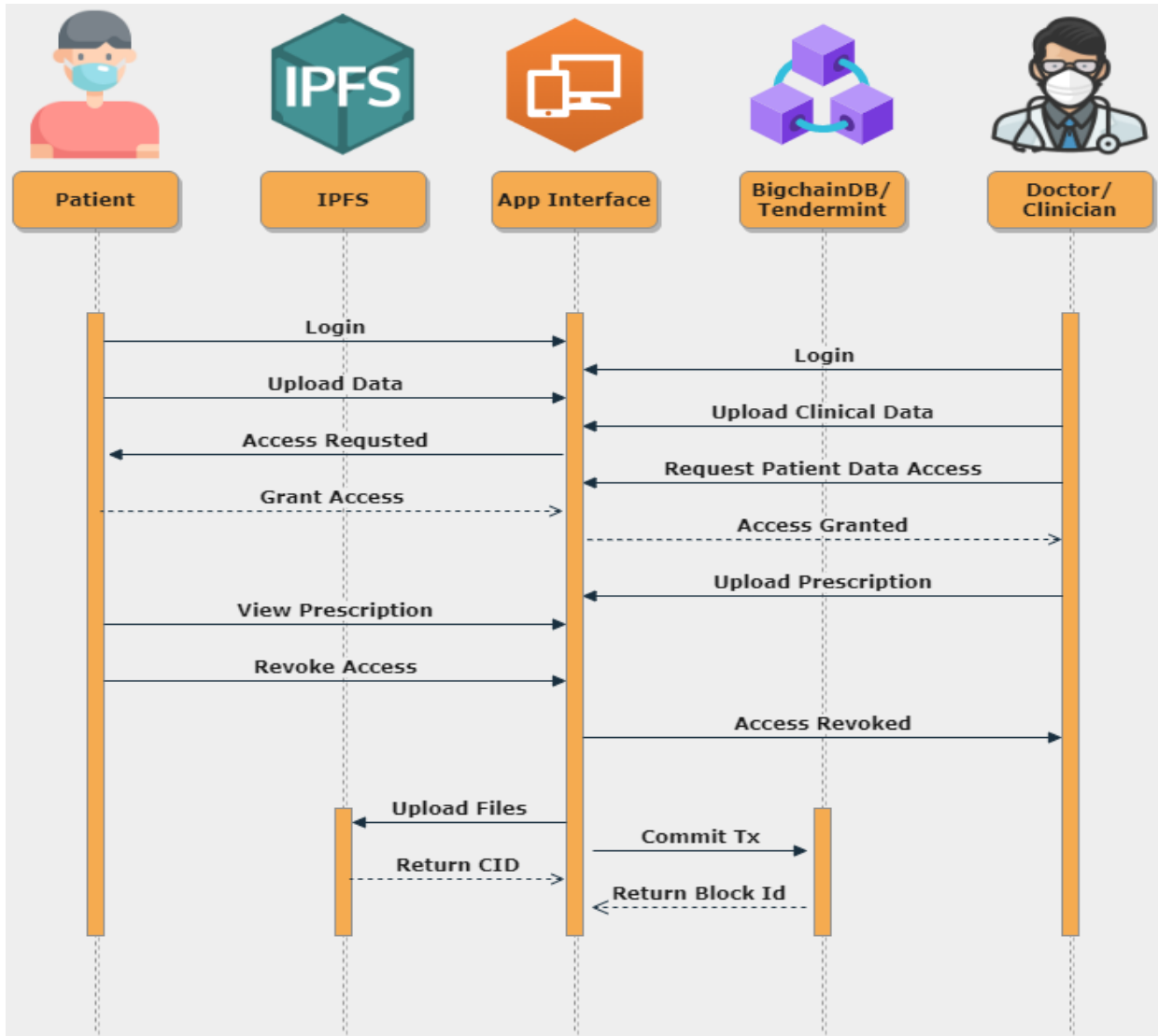

Fig.16 Basic overview of data exchange

### 6.2 Result and Discussion

The system is used for four main purposes: uploading health data, granting permission to the Doctor, accessing the patient data, and revoking the access. Another benefit is that the system offers RSA and blockchain-based security, bigchainDB allows users to query their data. Current systems utilize cloud technology to get information from a centralized database. A distributed ledger technology (DLT) runs our system, even if a node is damaged. However, if someone tries to alter the information, we can identify and correct that erroneous information. However, our system saves data in an immutable manner, so it will recover information even if the system crashes and loss of local data.

User Registration: Users can register by providing the appropriate detail. The doctor registration process has three steps (See Figure 17).

- Enter basic details First Name, Last Name, Email Address, Password, and Phone No.
- Verify using OTP received at the email address provided earlier.
- Provide professional details of Hospital, Qualification, Specialization, Work Experience, and Current Workspace.

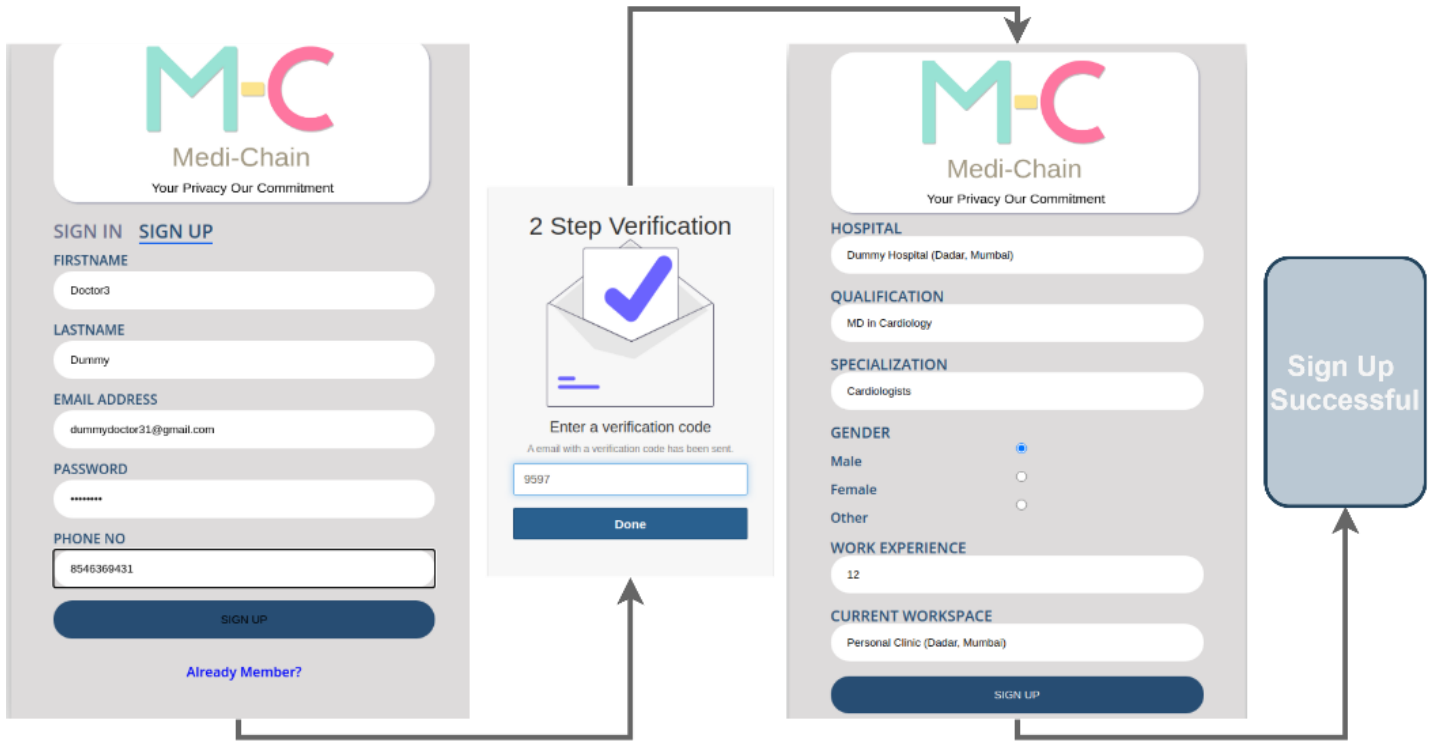

Fig. 17 Doctor Sign Up process

The patient registration process has two steps (See Figure 18).
- Enter basic details First Name, Last Name, Email Address, Password, Gender, Date of Birth, Phone No. and Emergency Email.
- Verify using OTP received at the email address provided earlier.

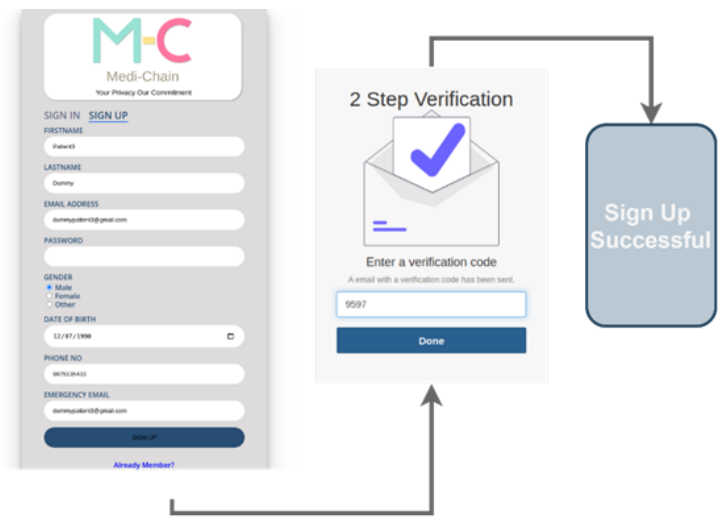

Fig. 18 Patient Sign Up Process

**Patient Portal:** After successful registration to the portal. Patients can now log in and interact with the features provided by the developed app like adding previous medical records, various lab records, radiology scans etc.

Initially, the Dashboard of doctors and other medical professionals are empty. Then Patients grants access to their resources (See Figure 19). After getting access to patient data Doctors prescribe medicine and other things accordingly (Figure 20). Then the patient can now remove access from the Doctor if they want (Figure 21).

Fig. 19 Granting the access to the Doctor

Fig. 20 Provide the prescription to the patient

Fig. 21 Revoking the access from the Doctor

After committing all these transactions on the blockchain. We can view all the blocks created on the blockchain using BigchainDb Dashboard in Figure 22, and transfer assets to the Doctor.

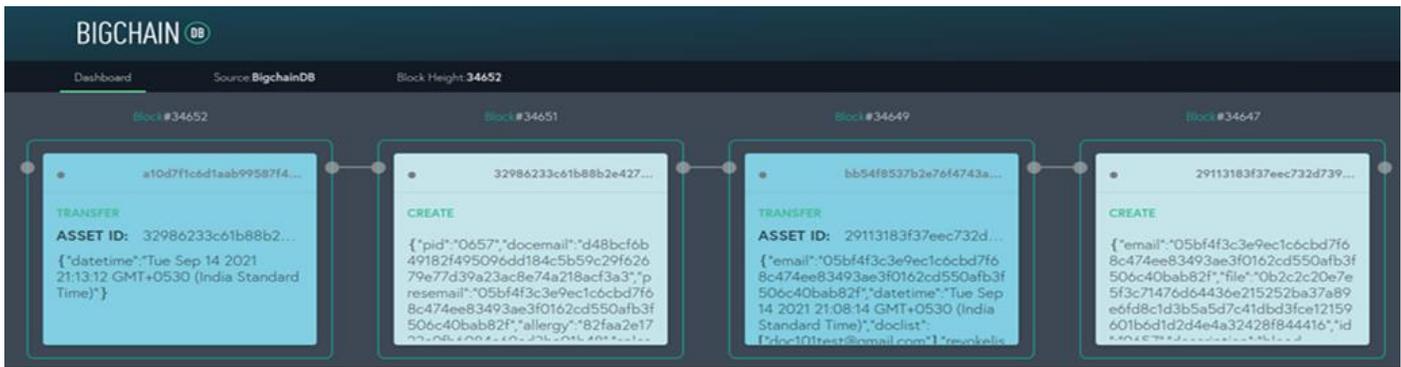

Fig. 22 Blockchain Dashboard

## 7. CONCLUSION AND FUTURE WORKS

Healthcare is one of the trending areas in the field of computer science after the adoption of advanced technology such as IoT, edge computing, and other AI-enabled technologies, but the existing healthcare systems facing several challenges such as non-patient centric systems, transparency, data breaches, privacy, and security of the patient data, etc. In addition, Current EHR systems suffer from significant shortcomings owing to the existence of potential data fragmentation, falsification of data, and system glitches that prevent access to critical information in the event of an emergency. This conventional EHR system is vulnerable to newer technologies that have been put in place to fight the current threats. Thus, there is a dire need to overcome the mentioned issues in the Healthcare domain. Hence, in this paper, a blockchain-assisted secure and reliable data exchange architecture for Cyber Physical Healthcare 4.0 is proposed. Initially, a system is proposed where electronic health records are stored and shared on a decentralized platform that can be utilized to build decentralized access using blockchain technology and replace the existing centralized system. Secondly, the proposed Blockchain-enabled architecture used the tendermint and IPFS technologies to resolve the issues like data fragmentation, data leakage, and illegal access to patient data that are prevalent in existing Electronic Health Record (EHR) systems. Finally, privacy and security are effectively preserved by considering the Blockchain-enabled AES-256 algorithm for data security. Thus, the proposed work provides a secure Cyber physical blockchain proofs-of-concept platform and a cost-effective solution. The proposed work may be extended to other applications by replacing more secure encryption techniques [37]. Moreover, the proposed work can be extended to transfer a large amount of data over the Cloud and monitor by intelligent technologies [38, 40]. In addition, ChatGPT and IoT can be combined with a cyber-physical healthcare system to expedite and improve patient care [39]. They make an outstanding duo that is changing how people interact with technology and may greatly improve our lives in the decades to come [41].

**Declaration of Competing Interest Statement**

The authors declare that no conflict of interest and they have no known competing financial interests or personal relationships that could have appeared to influence the work reported in this paper.